\documentclass[runningheads]{llncs}
\usepackage{times}
\usepackage{amssymb}
\usepackage[bookmarks=false]{hyperref}
\makeatletter
\def\@citex[#1]#2{\if@filesw\immediate\write\@auxout{\string\citation{#2}}\fi
  \@tempcnta\z@\@tempcntb\m@ne\def\@citea{}\@cite{\@for\@citeb:=#2\do
    {\@ifundefined
       {b@\@citeb}{\@citeo\@tempcntb\m@ne\@citea\def\@citea{,}{\bfseries
        ?}\@warning
       {Citation `\@citeb' on page \thepage \space undefined}}%
    {\setbox\z@\hbox{\global\@tempcntc0\csname b@\@citeb\endcsname\relax}%
     \ifnum\@tempcntc=\z@ \@citeo\@tempcntb\m@ne
       \@citea\def\@citea{,\,}\hbox{\csname b@\@citeb\endcsname}%
     \else
      \advance\@tempcntb\@ne
      \ifnum\@tempcntb=\@tempcntc
      \else\advance\@tempcntb\m@ne\@citeo
      \@tempcnta\@tempcntc\@tempcntb\@tempcntc\fi\fi}}\@citeo}{#1}}
\long\def\@makecaption#1#2{%
  \small
  \vskip\abovecaptionskip
  \sbox\@tempboxa{{\bfseries #1.} #2}%
  \ifdim \wd\@tempboxa >\hsize
    {\bfseries #1.} #2\par
  \else
    \global \@minipagefalse
    \hb@xt@\hsize{\hfil\box\@tempboxa\hfil}%
  \fi
  \vskip\belowcaptionskip}
\makeatother
\newcommand{\mc}[1]{\mathcal{#1}}
\newcommand{\nil}{\mathbf{0}}
\newcommand{\mv}[1]{\mathrel{\stackrel{#1}{\rightarrow}}}
\newcommand{\var}[1][]{\ensuremath{\mathit{var}_{#1}}}
\newcommand{\wfpreo}{\leq_{\rm WF}}
\newcommand{\norm}{\ensuremath{\mathit{norm}}}
\def\lastandname{~\&}
\spnewtheorem{observation}{Observation}{\bfseries}{\itshape}

\begin{document}

\mainmatter

\title{On Finite Bases for Weak Semantics:\\ Failures versus Impossible
Futures\thanks{This work is partially supported by the Dutch Bsik
project BRICKS.}}

\author{Taolue Chen \inst{1} \and Wan Fokkink \inst{1,2} \and Rob van Glabbeek\inst{3,4}}
\institute{CWI, Department of Software Engineering, PO Box 94079,
1090 GB Amsterdam,\\ The Netherlands   \and
Vrije Universiteit Amsterdam, Department of Theoretical Computer Science,\\
De Boelelaan 1081a, 1081 HV Amsterdam, The Netherlands \and
National ICT Australia, Locked Bag 6016, Sydney, NSW 1466,
Australia \and The University of New South Wales, School of
Computer Science and Engineering,\\ Sydney, NSW 2052, Australia}

\maketitle
\setcounter{footnote}{0}

\begin{abstract}
  We provide a finite basis for the (in)equational theory of the
  process algebra BCCS modulo the weak failures preorder and
  equivalence. We also give positive and negative results regarding
  the axiomatizability of BCCS modulo weak impossible futures semantics.
\end{abstract}

\section{Introduction}

Labeled transition systems constitute a widely used model of
concurrent computation. They model processes by explicitly
describing their states and their transitions from state to state,
together with the actions that produce these transitions. Several
notions of behavioral semantics have been proposed, with the aim
to identify those states that afford the same observations
\cite{Gla01,Gla93}. For equational reasoning about processes, one
needs to find an axiomatization that is sound and
\emph{ground-complete} modulo the semantics under consideration,
meaning that all equivalent closed terms can be equated. Ideally,
such an axiomatization is also \emph{$\omega$-complete}, meaning
that all equivalent \emph{open} terms can be equated. If such a
finite axiomatization exists, it is said that there is a
\emph{finite basis} for the equational theory.

For concrete semantics, so in the absence of the silent action
$\tau$, the existence of finite bases is well-studied
\cite{Gro90,Gla01,CFLN07}, in the context of the process algebra
BCCSP, containing the basic process algebraic operators from CCS
and CSP\@. However, for weak semantics, that take into account the
$\tau$, hardly anything is known on finite bases. In \cite{Gla93},
Van Glabbeek presented a spectrum of weak semantics. For several
of the semantics in this spectrum, a sound and ground-complete
axiomatization has been given, in the setting of the process
algebra BCCS (BCCSP extended by $\tau$), see, e.g., \cite{Gla97}.
But a finite basis has been given only for \emph{weak}, \emph{delay},
$\eta$- and \emph{branching bisimulation} semantics \cite{Mi89a,vG93a},
and in case of an infinite alphabet of actions also for \emph{weak
impossible futures} semantics \cite{VM01}. The reason for this lack of
results on finite bases, apart from the inherent difficulties arising
with weak semantics, may be that it is usually not so straightforward
to define a notion of unique normal form for \emph{open} terms in a
\emph{weak} semantics. Here we will employ a saturation technique,
in which normal forms are saturated with subterms.

In this paper, we focus on two closely related weak semantics,
based on failures and impossible futures. A \emph{weak failure}
consists of a trace $a_1\cdots a_n$ and a set $A$, both of
concrete actions. A state exhibits this weak failure pair if it
can perform the trace $a_1\cdots a_n$ (possibly intertwined with
$\tau$'s) to a state that cannot perform any action in $A$ (even
after performing $\tau$'s). In a \emph{weak impossible future},
$A$ can be a set of traces. Weak failures semantics plays an
essential role for the process algebra CSP \cite{BHR84}. For
convergent processes, it coincides with testing semantics
\cite{DeNHen84,RenVog07}, and thus is the coarsest congruence for
the CCS parallel composition that respects deadlock behavior.
Weak impossible futures semantics \cite{Vog92} is a natural
variant of possible futures semantics \cite{RB81}. In \cite{GV06}
it is shown that weak impossible futures semantics, with an
additional root condition, is the coarsest congruence containing
weak bisimilarity with explicit divergence that respects
deadlock/livelock traces (or fair testing, or any liveness
property under a global fairness assumption) and assigns unique
solutions to recursive equations.

The heart of our paper is a finite basis for the inequational theory
of BCCS modulo the weak failures \emph{preorder}. The axiomatization
consists of the standard axioms A1-4 for bisimulation, three extra
axioms WF1-3 for failures semantics, and in case of a finite alphabet
$A$, an extra axiom WF$_A$. The proof that A1-4 and WF1-3 are a finite
basis in case of an infinite alphabet is a sub-proof of the proof that
A1-4, \mbox{WF1-3} and WF$_A$ are a finite basis in case of a finite
alphabet. Our proof has the same general structure as the beautiful
proof for testing equivalences given in \cite{DeNHen84} and further
developed in \cite{Hen88}. Pivotal to this is the construction of
``saturated'' sets of actions within a term \cite{DeNHen84}.  Since
here we want to obtain an $\omega$-completeness result, we extend this
notion to variables.  Moreover, to deal with $\omega$-completeness, we
adopt the same general proof structure as in the strong case
\cite{FN05}. In this sense, our proof strategy can be viewed as a
combination of the strategies proposed in \cite{DeNHen84} and
\cite{FN05}.  Furthermore, we apply an algorithm from
\cite{AFI07,FG07,CFG08b} to obtain a finite basis for BCCS modulo weak
failures \emph{equivalence} for free.

At the end, we investigate the equational theory of BCCS modulo
weak impossible futures semantics. This shows a remarkable
difference with weak failures semantics, in spite of the strong
similarity between the definitions of these semantics (and between
their ground-complete axiomatizations). As said, in case of an
infinite alphabet, BCCS modulo the weak impossible futures
preorder has a finite basis \cite{VM01}. However, we show that in
case of a finite alphabet, such a finite basis does not exist.
Moreover, in case of weak impossible futures \emph{equivalence},
there is no finite ground-complete axiomatization, regardless of the
cardinality of the alphabet.

A finite basis for the equational theory of BCCSP modulo
(concrete) failures semantics was given in \cite{FN05}. The
equational theory of BCCSP modulo (concrete) impossible futures
semantics is studied in \cite{CF08}. It is interesting to see that
our results for weak semantics agree with their concrete
counterparts, with very similar proofs. This raises a challenging open
question: can one establish a general theorem to link the
axiomatizability (or nonaxiomatizability) of concrete and weak semantics?

An extended abstract of this paper appears as \cite{CFG08}.
\newpage

\section{Preliminaries} \label{sec2}

${\rm BCCS}(A)$ is a basic process algebra for expressing finite
process behavior. Its signature consists of the constant $\nil$,
the binary operator $\_+\_$\,, and unary prefix operators $\tau\_$
and $a\_$\,, where $a$ is taken from a nonempty set $A$ of visible
actions, called the \emph{alphabet}, ranged over by $a,b,c$. We
assume that $\tau\notin A$ and write $A_\tau$ for $A\cup\{\tau\}$,
ranged over by $\alpha,\beta$.
\[ t::= 0 \mid at \mid \tau t \mid t+t \mid x\]
Closed ${\rm BCCS}(A)$ terms, ranged over by $p,q$, represent
finite process behaviors, where $\nil$ does not exhibit any
behavior, $p+q$ offers a choice between the behaviors of $p$ and
$q$, and $\alpha p$ executes action $\alpha$ to transform into
$p$.  This intuition is captured by the transition rules below.
They give rise to $A_\tau$-labeled transitions between closed BCCS terms.
\[
  \frac{~}{\alpha x\mv{\alpha}x}
\qquad
  \frac{x\mv{\alpha}x'}{x+y\mv \alpha x'}
\qquad
  \frac{y\mv{\alpha}y'}{x+y\mv \alpha y'}
\]
We assume a countably infinite set $V$ of variables; $x,y,z$
denote elements of $V$. Open BCCS terms, denoted by $t,u,v,w$, may
contain variables from $V$. Write $\var(t)$ for the set of variables
occurring in $t$.
The operational semantics is extended verbatim to open terms;
variables generate no transition.
We write $t\Rightarrow u$ if there is a sequence of
$\tau$-transitions $t\mv{\tau}\cdots\mv{\tau}u$; furthermore
$t\mv{\alpha}$ denotes that there is a term $u$ with
$t\mv{\alpha}u$, and likewise $t\Rightarrow\mv{\alpha}$ denotes that
there are a terms $u,v$ with $t\Rightarrow u \mv{\alpha} v$.

The \emph{depth} of a term $t$, denoted by $|t|$, is the length of
the \emph{longest} trace of $t$, not counting $\tau$-transitions.
It is defined inductively as follows: $|\nil|=|x|=0$;
$|at|=1+|t|$; $|\tau t| = |t|$; $|t+u|=\max\{|t|, |u|\}$.

A (closed) substitution, ranged over by $\sigma,\rho$, maps variables
in $V$ to (closed) terms. For open terms $t$ and $u$, and a
preorder $\sqsubseteq$ (or equivalence $\equiv$) on closed terms, we
define $t\sqsubseteq  u$ (or $t\equiv u$) if $\sigma(t)\sqsubseteq\sigma(u)$
(resp.\ $\sigma(t)\equiv\sigma(u)$) for all closed substitutions
$\sigma$. Clearly, $t\mv{a}t'$ implies that
$\sigma(t)\mv{a}\sigma(t')$ for all substitutions $\sigma$.

An \emph{axiomatization} is a collection of equations $t \approx
u$ or of inequations $t \preccurlyeq u$. The (in)equations in an
axiomatization $E$ are referred to as \emph{axioms}.  If $E$ is an
equational axiomatization, we write $E\vdash t\approx u$ if the
equation $t\approx u$ is derivable from the axioms in $E$ using
the rules of equational logic (reflexivity, symmetry,
transitivity, substitution, and closure under BCCS contexts). For
the derivation of an inequation $t\preccurlyeq u$ from an
inequational axiomatization $E$, denoted by $E\vdash t\preccurlyeq
u$, the rule for symmetry is omitted. We will also allow equations
$t\approx u$ in inequational axiomatizations, as an abbreviation
of $t\preccurlyeq u\land u\preccurlyeq t$.

An axiomatization $E$ is \emph{sound} modulo a preorder $\sqsubseteq$
(or equivalence $\equiv$) if for all terms $t,u$, from $E\vdash
t\preccurlyeq u$ (or $E\vdash t\approx u$) it follows that
$t\sqsubseteq u$ (or $t\equiv u$).  $E$ is \emph{ground-complete} for
$\sqsubseteq$ (or $\equiv$) if $p\sqsubseteq q$ (or $p\equiv q$)
implies $E\vdash p\preccurlyeq q$ (or $E\vdash p\approx q$) for all
closed terms $p,q$. Moreover, $E$ is \emph{$\omega$-complete} if for
all terms $t,u$ with $E\vdash\sigma(t)\preccurlyeq\sigma(u)$ (or
$E\vdash\sigma(t)\approx\sigma(u)$) for all closed substitutions
$\sigma$, we have $E\vdash t\preccurlyeq u$ (or $E\vdash t\approx u$).
When $E$ is $\omega$-complete as well as ground-complete, it is
\emph{complete} for $\sqsubseteq$ (or $\equiv$) in the sense that
$t\sqsubseteq u$ (or $t\equiv u$) implies $E\vdash t\preccurlyeq u$
(or $E\vdash t\approx u$) for all terms $t,u$.  The equational
theory of BCCS modulo a preorder $\sqsubseteq$ (or equivalence
$\equiv$) is said to be \emph{finitely based} if there exists a
finite, $\omega$-complete axiomatization that is sound and
ground-complete for BCCS modulo $\sqsubseteq$ (or $\equiv$).

A1-4 below are the core axioms for BCCS modulo bisimulation
semantics. We write $t=u$ if $\mbox{A1-4}\vdash t\approx u$.
{\small\[
\begin{array}{l@{\qquad}rcl}
{\rm A}1&x+y &~\approx~& y+x\\
{\rm A}2&(x+y)+z &~\approx~& x+(y+z)\\
{\rm A}3&x+x &~\approx~& x\\
{\rm A}4&x+\nil &~\approx~& x\\
\end{array}
\]}%
Summation $\sum_{i\in\{1,\ldots,n\}}t_i$ denotes $t_1+\cdots+t_n$,
where summation over the empty set denotes $\nil$. As binding
convention, $\_+\_$ and summation bind weaker than $\alpha\_$\,.
For every term $t$ there exists a finite set $\{\alpha_i t_i\mid
i\in I\}$ of terms and a finite set $Y$ of variables such that
$t=\sum_{i\in I} \alpha_i t_i + \sum_{y\in Y}y$. The $\alpha_i
t_i$ for $i\in I$ and the $y\in Y$ are called the \emph{summands}
of $t$.  For a set of variables $Y$, we will often denote the term
$\sum_{y\in Y}y$ by $Y$.

\begin{definition}[Initial actions]\rm
For any term $t$, the set $\mc{I}(t)$ of initial actions is
defined as $\mc{I}(t)=\{a\in A\mid t\Rightarrow \mv{a}\}$.
\end{definition}

\begin{definition}[Weak failures]\rm \label{def:weak-failures}
\vspace{-1ex}
\begin{itemize}
  \item A pair $(a_1\cdots a_k,B)$, with $k\geq 0$ and $B \subseteq A$,
  is a \emph{weak failure pair} of a process $p_0$ if there is a path
  $p_0\Rightarrow \mv{a_1} \Rightarrow \cdots \Rightarrow \mv{a_k} \Rightarrow p_k$
  with $\mc{I}(p_k) \cap B = \emptyset$.

  \item Write $p\wfpreo q$ if the weak failure pairs of $p$ are
  also weak failure pairs of $q$.

  \item The \emph{weak failures preorder} $\sqsubseteq_{\rm WF}$ is
  given by\\ $p\sqsubseteq_{\rm WF} q$ iff (1) $p\wfpreo q$ and
        (2) $p\mv{\tau}$ implies that $q\mv{\tau}$.

  \item \emph{Weak failures equivalence} $\equiv_{\rm WF}$ is defined
  as $\sqsubseteq_{\rm WF} \cap \sqsubseteq_{\rm WF}^{-1}$.
\end{itemize}
\end{definition}
It is well-known that $p\wfpreo q$ is \emph{not} a
\emph{precongruence} for BCCS: e.g., $\tau\nil\wfpreo\nil$
but $\tau\nil+a\nil \not\wfpreo \nil+a\nil$. However,
$\sqsubseteq_{\rm WF}$ is, meaning that $p_1\sqsubseteq_{\rm WF} q_1$
and $p_2\sqsubseteq_{\rm WF} q_2$ implies $p_1+p_2\sqsubseteq_{\rm WF}
q_1+q_2$ and $\alpha p_1\sqsubseteq_{\rm WF} \alpha q_1$ for
$\alpha\in A_\tau$. In fact, $\sqsubseteq_{\rm WF}$ is the coarsest
precongruence contained in $\wfpreo$.
Likewise, $\equiv_{\rm WF}$ is a \emph{congruence} for BCCS.

\section{A Finite Basis for Weak Failures Semantics} \label{sec3}

\subsection{Axioms for the Weak Failures Preorder}

On BCCS processes, the weak failures preorder as defined above
coincides with the inverse of the must-testing preorder of \cite{DeNHen84}.
A sound and ground-complete axiomatization of the must-testing preorder
preorder has been given in \cite{DeNHen84}, in terms of a language
richer than BCCS\@. After restriction to BCCS processes, and reversing
the axioms, it consists of A1-4 together with the axioms: {\small
$$\begin{array}{l@{\qquad}rcl}
\mbox{N1} & \alpha x + \alpha y &\approx& \alpha (\tau x + \tau y)\\
\mbox{N2} & \tau(x+y) &\preccurlyeq& x + \tau y\\
\mbox{N3} & \alpha x + \tau(\alpha y + z) &\approx& \tau(\alpha x + \alpha y + z)\\
\mbox{E1} & x &\preccurlyeq& \tau x + \tau y
\end{array}$$}%
Here we simplify this axiomatization to A1-4 and WF1-3 from Tab.~\ref{tab1}.
In fact it is an easy exercise to derive WF1-3 from N1, N2 and E1,
and N1, N2 and E1 from WF1-3. It is a little harder to check that N3
is derivable from the other three axioms (cf.~Lem.~\ref{lemma5}).

\begin{table}[t] \center
 \begin{tabular}{l@{\qquad}rcl}
 WF1   & $a x+ay$  & $\approx$ & $a(\tau x+\tau y)$\\
 WF2   & $\tau(x+y)$  & $\preccurlyeq$ & $\tau x+ y$\\
 WF3   & $x$  & $\preccurlyeq $ & $ \tau x+  y$\medskip\\
\end{tabular}\caption{Axiomatization for the weak failures preorder} \label{tab1}
\vspace{-5mm}
\end{table}

\begin{theorem} \label{theo:ground-complete}
\mbox{\rm A1-4+WF1-3} is sound and ground-complete for
$\mathrm{BCCS}(A)$ modulo $\sqsubseteq_{\rm WF}$.
\end{theorem}
In this section, we extend this ground-completeness result with
two $\omega$-completeness results. The first one says, in
combination with Theo.~\ref{theo:ground-complete}, that as long
as our alphabet of actions is infinite, the axioms A1-4+WF1-3
constitute a finite basis for the inequational theory of BCCS$(A)$
modulo $\sqsubseteq_{\rm WF}$.

\begin{theorem} \label{theo:A-infinite}
If $|A|\mathbin=\infty$, then \mbox{\rm A1-4+WF1-3} is $\omega$-complete for
$\mathrm{BCCS}(A)$ modulo $\sqsubseteq_{\rm WF}$.
\end{theorem}
To get a finite basis for the inequational theory of BCCS modulo
$\sqsubseteq_{\rm WF}$ in case $|A|<\infty$, we need to add the
following axiom:
\[\mathrm{WF}_A\qquad \sum_{a\in A} ax_a \preccurlyeq \sum_{a\in A}a x_a +y\]
where the $x_a$ for $a\in A$ and $y$ are distinct variables.

\begin{theorem} \label{theo:A-finite}
If $|A|<\infty$, then \mbox{\rm A1-4+WF1-3+WF}$_A$ is $\omega$-complete for
 $\mathrm{BCCS}(A)$ modulo $\sqsubseteq_{\rm WF}$.
\end{theorem}
The rest of this section up to Sec.~\ref{sec:wfe} is devoted to the
proofs of Theorems~\ref{theo:ground-complete}--\ref{theo:A-finite}.
For a start, the inequations in Tab.~\ref{tab:D-eqs} can be derived
from A1-4+WF1-3:

\vspace{-3mm}
\begin{table}
\begin{center}
 \begin{tabular}{l@{\qquad}rcl}
 D1 & $ \tau (x+y)+x$  & $\approx$ & $\tau (x+y)$\\
 D2 & $\tau(\tau x+y)$  & $\approx$ & $\tau x+ y$\\
 D3 & $ ax + \tau(ay+z)$  & $\approx$ & $\tau(a x+ a y+z)$\\
 D4 & $ \tau x $  & $\preccurlyeq$ & $\tau x +y$\\
 D5 & $ \sum_{i\in I} a x_i $  & $\approx$ & $a (\sum_{i\in I}\tau
 x_i)\mbox{ for finite nonempty index sets }I $\\
 D6 & $ \tau x + y$  & $\approx$ & $\tau x + \tau (x+y) $\\
 D7 & $ \tau x + \tau y$  & $\approx$ & $\tau x + \tau (x+y) + \tau y $\\
 D8 & $ \tau x + \tau (x+y+z)$  & $\approx$ & $\tau x + \tau (x+y) + \tau(x+y+z) $\\
 D9 & $ \sum_{i\in I}\tau (at_i+y_i)$ & $\approx$ & $\sum_{i\in I}\tau (at + y_i)$
      for finite $I$, where $t = \sum_{i\in I} \tau t_i$.
\end{tabular}
\end{center}
\caption{Derived inequations} \label{tab:D-eqs}
\end{table}
\vspace{-7mm}
\begin{lemma} \label{lemma5}
  \mbox{\rm D1-9} are derivable from \mbox{\rm A1-4+WF1-3}.
\end{lemma}

\begin{proof}
We shorten ``$\mbox{A1-4+WF1-3} \vdash$'' to ``$\vdash$''.\vspace{-1ex}
\begin{enumerate}
  \item
 By WF3, $\vdash x \preccurlyeq \tau x$, and thus $\vdash\tau x+x\preccurlyeq
 \tau x$. Moreover, by WF2,\\ $\vdash\tau (x+x) \preccurlyeq \tau x+x$,
 hence $\vdash\tau x\preccurlyeq \tau x+x$. In summary, $\vdash\tau x\approx
 \tau x+x$.

 So $\vdash \tau(x+y)\approx \tau(x+y)+x+y+x \approx \tau(x+y)+x$.

 \medskip

  \item
 By WF2, $\vdash \tau(x+\tau x)\preccurlyeq \tau x+\tau x = \tau
 x$, so by D1, $\vdash \tau\tau x\preccurlyeq \tau x$.
 Hence, by WF2, $\vdash\tau (\tau x+y) \preccurlyeq \tau \tau x+y \preccurlyeq \tau x+y$.

 Moreover, by WF3, $\vdash\tau  x+y \preccurlyeq \tau(\tau x+y)$.

 \medskip

 \item By WF3, $\vdash y \preccurlyeq \tau y+\tau x$. So by WF1, $\vdash ay\preccurlyeq a(\tau x+\tau y)\approx ax+ay$. This implies $\vdash \tau(ay+z)\preccurlyeq \tau(ax+ay+z)$. Hence, by D1, $\vdash ax+\tau(ay+z)\preccurlyeq ax+ \tau(ax+ay+z) \approx \tau(ax+ay+z)$.

Moreover, by WF2, $ \vdash \tau(a x+ a y+z) \preccurlyeq ax +
\tau(ay+z)$.

\medskip
  \item By WF3 and D2, $\vdash \tau x\preccurlyeq \tau\tau x+y \approx \tau x+y$.
  \medskip

  \item By induction on $|I|$, using WF1 and D2.
  \medskip

  \item By D4 and D1, $\vdash \tau x+y \preccurlyeq \tau x+\tau(x+y)+y \approx \tau x+\tau(x+y)$.

        Moreover, by WF2, $\vdash \tau x+\tau(x+y)\preccurlyeq \tau x+\tau x+y=\tau x+y$.
  \medskip

  \item By D4 in one direction; by D6 and D1 in the other.
  \medskip

  \item By D4 in one direction; by D6 and D1 in the other.
  \medskip

  \item By D1, $\vdash \sum_{i\in I}\tau (at_i+y_i) \approx
        \sum_{i\in I}\tau (at_i+y_i) +u$, where $u=\sum_{i\in I}at_i$.
    Thus, by repeated application of D3, $\vdash \sum_{i\in I}\tau (at_i+y_i) \approx
        \sum_{i\in I}\tau (at_i+u+y_i) = \sum_{i\in I}\tau (u+y_i)$.
    By D5 we have $u=at$.
\qed
\end{enumerate}
\end{proof}

\subsection{Normal Forms}

The notion of a normal form, which is formulated in the following
two definitions, will play a key role in the forthcoming proofs.
For any set $L\subseteq A\cup V$ of actions and variables let $A_L
= L \cap A$, the set of actions in $L$, and $V_L = L \cap V$, the
set of variables in $L$.

\begin{definition}[Saturated family]\rm Suppose $\mc{L}$ is a finite family of
finite sets of {actions} and {variables}. We say $\mc{L}$ is
\emph{saturated} if it is nonempty and
\begin{itemize}
  \item $L_1, L_2\in \mc{L}$ implies that $L_1\cup L_2 \in
  \mc{L}$; and

  \item $L_1, L_2\in \mc{L}$ and $L_1\subseteq L_3 \subseteq L_2$
  imply that $L_3\in \mc{L}$.
\end{itemize}
\end{definition}

\begin{definition}[Normal form]\rm
\begin{list}{$\bullet$}{\leftmargin 18pt
                        \labelwidth\leftmargini\advance\labelwidth-\labelsep}
  \item[(i)] A term $t$ is in $\tau$ normal form if\vspace{-6pt}
  \[t = \sum_{L\in \mc{L}}\tau \left(\sum_{a \in A_L} at_a+ V_L\right)\vspace{-3pt}\]
where the $t_a$ are in normal form and $\mc{L}$ is a saturated
family of sets of actions and variables. We write $L(t)$ for
$\bigcup_{L\in\mc{L}}L$; note that $L(t) \in \mc{L}$.
\medskip
  \item[(ii)] $t$ is in action normal form if
\[t= \sum_{a\in A_L} at_a + V_L \]
where the $t_a$ are in normal form and $L\subseteq A \cup V$. We
write $L(t)$ for $L$.
\medskip
\item[(iii)] $t$ is in normal form if it is either in $\tau$
normal form or in action normal form.\pagebreak[3]
\end{list}
\end{definition}
Note that the definition of a normal form requires that for any $a\in A$, if
$t\Rightarrow\mv{a} t_1$ and $t\Rightarrow\mv{a} t_2$, then $t_1$ and $t_2$ are
syntactically identical.

We prove that every term can be equated to a normal form. We start
with an example.

\begin{example}
 Suppose $t= \tau(at_1+\tau(bt_2+ct_3)+x) +\tau(at_4+\tau x+\tau y)+z$.
 Then $t$ can be equated to a $\tau$ normal form with
 $\mc{L}=\{${\small$\{a,b,c, x\},\{a, x,y\}, \{a,b,c, x,y,z\}$,
 $\{a,b,c, x,y\},\hspace{-.6pt}\{a,b,c, x,z\},\hspace{-.6pt}
 \{a,b, x,y\},\hspace{-.5pt} \{a,c, x,y\},\hspace{-.5pt}
 \{a, x,y,z\},\hspace{-.6pt} \{a,b, x,y,z\},\hspace{-.6pt}
 \{a,c, x,y,z\}$}$\}$.
 We give a detailed derivation. By D2,
 \[
   \vdash t \approx \tau(at_1+bt_2+ct_3+x) +\tau(at_4+ x+ y)+z
 \]
 By D6,
 \[
   \vdash t \approx \tau(at_1+bt_2+ct_3+x) +\tau(at_4+x+y)+\tau(at_4+x+y+z)
 \]
 Let $u_a =  \tau t_1+\tau t_4$, $u_b=t_2$ and $u_c=t_3$. By D9,
 \[
   \vdash t \approx \tau(au_a+bu_b+cu_c+x)+\tau(au_a+x+y)+\tau(au_a+x+y+z)
 \]
By induction, $u_a$ can be brought into a normal form $t_a$, and
likewise for $u_b$ and $u_v$. So
 \[
   \vdash t \approx \tau(tu_a+bt_b+ct_c+x)+\tau(at_a+x+y)+\tau(at_a+x+y+z)
 \]
By D7,
 \[
   \vdash t \approx \begin{array}[t]{l}
                    \tau(at_a+bt_b+ct_c+x) +\tau(at_a+x+y)\\
            +\tau(at_a+x+y+z) + \tau(at_a+bt_b+ct_c+x+y+z)
            \end{array}
 \]
Finally, by D8,
 \[
   \vdash t \begin{array}[t]{l@{}}
  \approx \begin{array}[t]{@{}l@{}} \tau(at_a+bt_b+ct_c+x)
          +\tau(at_a+x+y)   +\tau(at_a+x+y+z)\\
          + \tau(at_a+bt_b+ct_c+x+y+z)
          + \tau(at_a+bt_b+ct_c+x+y)\\
          + \tau(at_a+bt_b+ct_c+x+z)
          + \tau(at_a+bt_b+x+y)
          + \tau(at_a+ct_c+x+y)\\
          + \tau(at_a+bt_b+x+y+z)
          + \tau(at_a+ct_c+x+y+z)
             \end{array}\\
   = \sum_{L\in \mc{L}}\tau(\sum_{a\in A_L} a t_a + V_L)
  \end{array}
 \]
\end{example}

\begin{lemma} \label{nf}
  For any term $t$, $\mbox{\rm A1-4+WF1-3} \vdash t\approx t'$ for some normal form $t'$.
\end{lemma}

\begin{proof}
By induction on $|t|$. We distinguish two cases.
\medskip
\begin{itemize}
 \item $t\,\not\!\mv{\tau}$. Let $t= \sum_{i\in I}a_i t_i+Y$. By D5,
\[\vdash t\approx \sum_{a\in \mc{I}(t)} a(\!\!\!\!\sum_{i\in I, a_i=a}\!\!\!\!\tau t_i)+Y~.\]
By induction, for each $a\in\mc{I}(t)$,
\[\vdash \!\!\!\!\sum_{i\in I,a_i=a}\!\!\!\! \tau t_i\approx t_a\]
for some normal form $t_a$. So we are done.
\medskip
 \item $t \mv{\tau}$. By D6, $t$ can be brought in the form
 $\sum_{i\in I}\tau t_i$ with $I\neq\emptyset$, and using D2 one can
 even make sure that $t_i\,\not\!\mv{\tau}$ for $i\in I$.
 Using the first case in this proof, we obtain,
 for each $i\in I$,
\[
 \vdash t_i \approx \sum_{a\in A_{L(i)}}a t_{a,i} + V_{L(i)}
\]
for some $L(i) \subseteq A \cup V$.
Thus\vspace{-3pt}
\[
\vdash t \approx \sum_{i\in I}
\tau \left(\sum_{a\in A_{L(i)}}a t_{a,i} + V_{L(i)}\right) ~.
\vspace{-3pt}\]
For each $a\in \mc{I}(t)$, we define \qquad
$\displaystyle u_a=\!\!\!\!\sum_{i\in I,~ a \in A_{L(i)}}\!\!\!\! \tau t_{a,i}~.$\\[1ex]
Then $|u_a| < |t|$.
By induction, $\vdash u_a \approx t_a$ for some normal form $t_a$.\\
Define $\mc{L} = \{L(i) \mid i \in I\}$.
By repeated application of D9 we obtain\vspace{-3pt}
\[
\vdash t \approx \sum_{i\in I} \tau \left( \sum_{a\in
  A_{L(i)}}\!\!a u_{a} + V_{L(i)}\right)
\approx \sum_{L\in \mc{L}}\tau\left( \sum_{a\in A_L}\ at_a +V_L\right)
~.\vspace{-3pt}\]
The latter term has the required form, except that the family $\mc{L}$
need not be saturated. However, it is straightforward to saturate
$\mc{L}$ by application of D7 and D8.
\qed
\end{itemize}
\end{proof}

\begin{lemma} \label{essential}
  Suppose $t$ and $u$ are both in normal forms and $t\sqsubseteq_{\rm WF} u$.
  If $t\Rightarrow \mv{a}t_a$, then there exists a term $u_a$ such that $u\Rightarrow\mv{a} u_a$ and $t_a\wfpreo u_a$.
\end{lemma}

\begin{proof}
 Suppose $t\sqsubseteq_{\rm WF} u$ and $t\Rightarrow \mv{a} t_a$.
 Let $\sigma$ be the closed substitution given by $\sigma(x)=\nil$ for
 all $x\in V$.  As $(a,\emptyset)$ is a weak failure
 pair of $\sigma(t)$ and $\sigma(t)\sqsubseteq_{\rm WF} \sigma(u)$, it
 is also a weak failure pair of $u$. Thus there
 exists a term $u_a$ such that
 \raisebox{0pt}[0pt][0pt]{$u\Rightarrow\mv{a} u_a$}.
 By the definition of a normal form, this term is unique.\hfill(*)

 We now show that $t_a \wfpreo u_a$.
 Let $\rho$ be a closed substitution.
 Consider a weak failure pair $(a_1\cdots a_k,B)$ of $\rho(t_a)$.
 Then $(aa_1\cdots a_k,B)$ is a weak failure pair of
 $\rho(t)$, and hence also of $\rho(u)$. It suffices to conclude that
 $(a_1\cdots a_k,B)$ is a weak failure pair of $\rho(u_a)$.
 However, we can \emph{not} conclude this directly, as
 possibly $u\Rightarrow x+u'$ where
 $(a a_1\cdots a_k,B)$ is a weak failure pair of $\rho(x)$.
 To ascertain that nevertheless $(a_1\cdots a_k,B)$ is a weak
 failure pair of $\rho(u_a)$,
 we define a modification $\rho'$ of $\rho$ such that
 for all $\ell\leq k$ and for all terms $v$, $\rho(v)$ and $\rho'(v)$
 have the same weak failure pairs $(c_1\cdots c_\ell,B)$, while for all
 $x\in V$, $(a a_1\cdots a_k,B)$ is not a weak failure pair of $\rho'(x)$.

 We obtain $\rho'(x)$ from $\rho(x)$ by replacing subterms $bp$ at
 depth $k$ by $\nil$ if $b\not\in B$ and by $bb\nil$ if $b\in B$.
 That is,
 \newcommand{\chop}[1][]{\mathit{chop}_{#1}}
 \[
   \rho'(x)=\chop[k](\rho(x))
 \]
 with ${\it chop}_m$ for all $m\geq 0$ inductively defined by
 \[ \begin{array}{lcl}
        \chop[m](\nil) &=& \nil \\
        \chop[m](p+q)  &=& \chop[m](p)+\chop[m](q) \\
        \chop[m](\tau p) &=& \tau\,\chop[m](p)\\
        \chop[0](bp)   &=& \left \{ \begin{array}{ll}
                              \nil   & \mbox{if $b\not\in B$} \\
                              bb\nil & \mbox{if $b\in B$}
                                    \end{array} \right. \\
        \chop[m+1](bp) &=& b\,\chop[m](p)
      \end{array}
 \]
 We proceed to prove that $\rho'$ has the desired properties
 mentioned above.

 \begin{enumerate}
 \renewcommand{\theenumi}{\Alph{enumi}}
 \item \label{rhomodA}
   For all $\ell\leq k$ and $c_1,\ldots,c_\ell\in A$ and for all terms
   $v$, $\rho(v)$ and
   $\rho'(v)$ have the same weak failure pairs $(c_1\cdots c_\ell,B)$,

   The difference between $\rho(v)$ and $\rho'(v)$ only appears within
   subterms of depth $k$, that is for terms $p$ such that
   $\rho(v)\Rightarrow \mv{c_1} \Rightarrow \cdots \Rightarrow
   \mv{c_k} \Rightarrow p$ for certain $c_1,\ldots,c_k \in A$. Such a
   subterm $p$ of $\rho(v)$ corresponds to a subterm $p'$ of
   $\rho'(v)$---still satisfying $\rho'(v)\Rightarrow \mv{c_1}
   \Rightarrow \cdots \Rightarrow \mv{c_k} \Rightarrow p'$---in which
   certain subterms $bq$ are replaced by $\nil$ if $b\not\in B$ and by
   $bb\nil$ if $b\in B$.  For such corresponding subterms $p$ and $p'$
   we have $\mc{I}(p)\cap B=\emptyset$ if and only if $\mc{I}(p')\cap
   B=\emptyset$.  From this the claim follows immediately.

   \medskip

 \item \label{rhomodB}
   For all $x\in V$,
   $(a a_1\cdots a_k,B)$ is not a weak failure pair of $\rho'(x)$.

   \medskip

   To this end we show that for all closed terms $p$, $\chop[m](p)$
   does not have any weak failure pair $(c_0\cdots c_{m},B)$
   with $c_0,\ldots,c_m\in A$.
   We apply induction on $m$.

   \smallskip

   \noindent
   \textit{Base case:} Since the summands of $\chop[0](p)$, when
   skipping over initial $\tau$-steps, are
   $bb\nil$ with $b\in \mc{I}(p)\cap B$, $\chop[0](p)$ does not
   have a weak failure pair $(c_0,B)$.

   \smallskip

   \noindent
   \textit{Induction step:} Let $m>0$. By induction, for closed terms $q$,
   $\chop[m-1](q)$ does not have weak failure pairs $(c_1\cdots c_{m},B)$.
   Since the transitions of $\chop[m](p)$ are
   $\chop[m](p)\mv{c}\chop[m-1](q)$ for $p\mv{c}q$, it
   follows that $\chop[m](p)$ does not have weak failure pairs
   $(c_0\cdots c_{m},B)$.
 \end{enumerate}

 \noindent
 Now, since $(a_1\cdots a_k,B)$ is a weak failure pair of $\rho(t_a)$,
 by property (\ref{rhomodA}) it is also a weak failure pair of $\rho'(t_a)$,
 Therefore $(a a_1\cdots a_k,B)$ is a weak failure pair of $\rho'(t)$,
 and hence also of $\rho'(u)$.
 Since according to property (\ref{rhomodB}) it is \emph{not} the case that
 $u\Rightarrow x+u'$ with $(a  a_1\cdots a_k,B)$ a weak failure pair
 of $\rho'(x)$, it must be the case that $u\Rightarrow \mv{a}u''$ such
 that $(a_1\cdots a_k,B)$ is a weak failure pair of $\rho'(u'')$.
 By (*), $u''=u_a$.
 Again by property (\ref{rhomodA}), $(a_1\cdots a_k,B)$ is a
 weak failure pair of $\rho(u_a)$.
\qed
\end{proof}

\subsection{$\omega$-Completeness Proof}

We are now in a position to prove Theo.~\ref{theo:A-infinite}
($\omega$-completeness in case of an infinite alphabet) and
Theo.~\ref{theo:A-finite} ($\omega$-completeness in case of a
finite alphabet), along with Theo.~\ref{theo:ground-complete}
(ground completeness). We will prove these three theorems in one
go. Namely, in the proof, two cases are distinguished; only in the
second case ($\mc{I}(t)=A$), in which the $A$ is guaranteed to be
finite, will the axiom WF$_A$ play a role.

\begin{proof}
Let $t\sqsubseteq_{\rm WF} u$. We need to show that $\vdash
t\preccurlyeq u$. We apply induction on $|t|+|u|$. By
Lem.~\ref{nf}, we can write $t$ and $u$ in normal form.

We first prove that $L(t)\subseteq L(u)$. Suppose this is not the
case. Then there exists some $a\in A_{L(t)}\setminus A_{L(u)}$ or
some $x\in V_{L(t)}\setminus V_{L(u)}$. In the first case, let
$\sigma$ be the closed substitution with $\sigma(z)=\nil$ for all
$z\in V$; we find that $(a,\emptyset)$ is a weak failure pair of
$\sigma(t)$ but not of $\sigma(u)$, which contradicts the fact that
$\sigma(t)\sqsubseteq_{\rm WF} \sigma(u)$. In the second case, pick some
$d>\max\{|t|,|u|\}$, and consider the closed substitution
$\sigma(x)=a^d\nil$ and $\sigma(z)=\nil$ for $z\neq x$. Then
$(a^d,\emptyset)$ is weak failure pair of $\sigma(t)$. However, it
can \emph{not} be a weak failure pair of $\sigma(u)$, again
contradicting $\sigma(t)\sqsubseteq_{\rm WF} \sigma(u)$.

We distinguish two cases, depending on whether $\mc{I}(t)=A$ or
not.

  \begin{enumerate}

  \item $\mc{I}(t) \neq A$.  We distinguish three cases.
Due to the condition that $t\mv{\tau}$ implies  $u\mv{\tau}$, it
cannot be the case that $t$ is an action normal form and $u$ a
$\tau$ normal form.

\medskip
  \begin{enumerate}
\item $t$ and $u$ are both action normal forms.
So $t= \sum_{a\in A_L} at_a + V_L$ and $u= \sum_{a\in A_M}
au_a+V_M$. We show that $L(t)=L(u)$. Namely, pick $b\in A\setminus
A_L$, and let $\sigma$ be the closed substitution with
$\sigma(z)=\nil$ for any $z\in V_L$, and $\sigma(z)=b\nil$ for
$z\not\in V_L$. As $(\varepsilon,A\setminus\mc{I}(t))$ is a weak
failure pair of $t$, and
  hence of $u$, it must be that $L(u)\subseteq L(t)$.
Together with $L(t)\subseteq L(u)$ this gives $L(t)=L(u)$. By
Lem.~\ref{essential}, for each $a\in \mc{I}(t)$, $t_a \wfpreo
u_a$, and thus clearly  $t_a\sqsubseteq_{\rm WF} \tau u_a$. By
induction, $\vdash t_a\preccurlyeq \tau u_a$ and hence $\vdash
at_a\preccurlyeq au_a$. It follows that
\[\vdash t = \sum_{a\in A_L} at_a +V_L \preccurlyeq \sum_{a\in A_L}
au_a +V_L  =  \sum_{a\in A_M} au_a +V_M = u\]

\item Both $t$ and $u$ are $\tau$ normal forms:
\[ t= \sum_{L\in \mc{L}}\tau (\sum_{a \in A_L} at_a+ V_L)\]
and
\[ u = \sum_{M\in \mc{M}}\tau (\sum_{a \in A_M} au_a+ V_M)\]

By Lem.~\ref{essential}, for each $a\in  \mc{I}(t)$, $t_a \wfpreo
u_a$, and thus clearly $t_a\sqsubseteq_{\rm WF} \tau u_a$. By
induction, $\vdash t_a\preccurlyeq \tau u_a$. By these
inequalities, together with D4,
\begin{equation} \label{eqnX}
\vdash t \preccurlyeq \sum_{L\in \mc{L}} \tau(\sum_{a\in A_L} au_a
+ V_L) + u
\end{equation}

\medskip

We now show that $\mc{L}\subseteq \mc{M}$. Take any $L\in \mc{L}$,
pick $b\in A \setminus A_L$, and consider the closed substitution
$\sigma(z)=\nil$ for any $z\in V_L$, and $\sigma(z)=b\nil$ for
$z\not\in V_L$. Since $\sigma(t)\mv{\tau} \sigma(\sum_{a\in L}at_a)$
and $\sigma(t)\sqsubseteq_{\rm WF} \sigma(u)$, there exists an $M\in
\mc{M}$ with $A_M\subseteq A_L$ and $V_M \subseteq V_L$. Since
also $L\subseteq L(t) \subseteq L(u)$, and $\mc{M}$ is saturated,
it follows that $L\in\mc{M}$. Hence, $\mc{L} \subseteq \mc{M}$.

\medskip

Since $\mc{L} \subseteq \mc{M}$,
\begin{equation} \label{eqnY}
\sum_{L\in \mc{L}}\tau(\sum_{a\in A_L} au_a + V_L) + u = u
\end{equation}
By (\ref{eqnX}) and (\ref{eqnY}), $ \vdash t\preccurlyeq u$.

\medskip

\item $t$ is an action normal form and $u$ is a $\tau$ normal
form. Then $\tau t \sqsubseteq_{\rm WF} u$. Note that $\tau t$ is a
$\tau$ normal form, so according to the previous case,
\[\vdash \tau t \preccurlyeq u\]
By WF3,
\[\vdash t\preccurlyeq \tau t \preccurlyeq u \]

\end{enumerate}

   \medskip

   \item $\mc{I}(t)=A$. Note that in this case, $|A|<\infty$. So, according to
Theo.~\ref{theo:A-finite}, axiom WF$_A$ is at our disposal. As
before, we distinguish three cases.
\medskip
   \begin{enumerate}
    \item Both $t$ and $u$ are action normal forms. Since
    $L(t)\subseteq L(u)$ we have
   $t= \sum_{a\in A} at_a +W$ and $u= \sum_{a\in A} au_a+X$ with $W
   \subseteq X$. By WF$_A$,
   \[ \vdash \sum_{a\in A}a t_a \preccurlyeq \sum_{a\in A}a t_a + u\]
   By Lem.~\ref{essential}, for each $a\in A$, $t_a \wfpreo u_a$, and
   thus clearly $t_a\sqsubseteq_{\rm WF} \tau u_a$. By induction,
   $\vdash t_a\preccurlyeq \tau u_a$. It follows, using $W\subseteq X$, that
  \[ \vdash t=\sum_{a\in A}a t_a +W \preccurlyeq \sum_{a\in A}a u_a + u + W=u\]

    \medskip

   \item Both $t$ and $u$ are $\tau$ normal forms.
          \[t= \sum_{L\in \mc{L}}\tau (\sum_{a \in A_L} at_a+V_L)\]
    and
    \[     u = \sum_{M\in \mc{M}}\tau (\sum_{a \in A_M} au_a+V_M)\]
    By D1 and WF$_A$ (clearly, in this case $A_{L(t)}=A$),
   \begin{equation} \label{eqPP}
   \vdash t\approx t+\sum_{a\in A}a t_a \preccurlyeq t+\sum_{a\in A}a t_a+u
   \end{equation}
   By Lem.~\ref{essential}, for each $a\in A$, $t_a \wfpreo u_a$, and thus clearly
   $t_a\sqsubseteq_{\rm WF} \tau u_a$. By induction, $\vdash
   t_a\preccurlyeq \tau u_a$. By these inequalities, together with (\ref{eqPP}),
   \[
   \vdash t \preccurlyeq \sum_{L\in \mc{L}}\tau (\sum_{a \in A_L} au_a+V_L)
   + \sum_{a\in A}a u_a + u
   \]
   So by D1,
   \begin{equation} \label{eqnQQ} \vdash t \preccurlyeq \sum_{L\in \mc{L}}
   \tau (\sum_{a \in A_L} au_a+V_L) + u\end{equation}
    Now for $L\in \mc{L}$ with $A_L\neq A$ we have $L \in \mc{M}$ using
    the same reasoning as in 1(b). For $L\in \mc{L}$ with $A_L=A$ we
    have $V_L \subseteq V_{L(t)} \subseteq V_{L(u)}$. By WF$_A$ we have
    \begin{equation} \label{eqnSS}
    \vdash \tau(\sum_{a \in A_L} au_a+V_L) \preccurlyeq
    \tau(\sum_{a \in A} au_a+V_{L(u)})
    \end{equation}
    As the latter is a summand of $u$ we obtain $t\preccurlyeq u$.

\medskip

  \item $t$ is an action normal form and $u$ is a $\tau$ normal
  form. This can be dealt with as in case 1(c).
   \end{enumerate}
  \end{enumerate}
  This completes the proof. \qed
\end{proof}

\subsection{Weak Failures Equivalence} \label{sec:wfe}

In \cite{AFI07,FG07} an algorithm is presented which takes as
input a sound and ground-complete inequational axiomatization $E$
for BCCSP modulo a preorder $\sqsubseteq$ which \emph{includes the
ready simulation preorder} and is \emph{initials
preserving},\footnote{meaning that $p\sqsubseteq q$ implies that
$I(p)\subseteq I(q)$, where the set $I(p)$ of \emph{strongly}
initial actions is $I(p) = \{\alpha \in A_{\tau} \mid
p\mv{\alpha}\}$} and generates as output an equational
axiomatization $\mc{A}(E)$ which is sound and ground-complete for
BCCSP modulo the corresponding equivalence---its kernel: $\sqsubseteq
\cap \sqsubseteq^{-1}$. Moreover, if the
original axiomatization $E$ is $\omega$-complete, so is the
resulting axiomatization. The axiomatization $\mc{A}(E)$ generated
by the algorithm from $E$ contains the axioms A1-4 for
bisimulation equivalence and the axioms $\beta(\alpha x + z) +
\beta (\alpha x + \alpha y + z) \approx
 \beta (\alpha x + \alpha y + z)$ for $\alpha,\beta\in A_\tau$
that are valid in ready simulation semantics, together with the
following equations, for each inequational axiom $t \preccurlyeq
u$ in $E$:
\begin{itemize}

\item $ t + u \approx u$; and

\item  $\alpha(t + x) + \alpha(u + x) \approx \alpha(u + x)$ (for
each $\alpha\in A_\tau$, and some variable $x$ that does not occur
in $t + u$).
\end{itemize}
Moreover, if $E$ contains an equation (formally abbreviating two
inequations), this equation is logically equivalent to the four
equations in $\mc{A}(E)$ that are derived from it, and hence can be
incorporated in the equational axiomatization unmodified.

Recently, we lifted this result to weak semantics
\cite{CFG08b}, which makes the aforementioned algorithm applicable
to all 87 preorders surveyed in \cite{Gla93} that are at least as
coarse as the ready simulation preorder. Namely, among others, we
show that

\begin{theorem}\label{alg}
Let $\sqsubseteq$ be a weak initials preserving precongruence%
\footnote{meaning that $p\sqsubseteq q$ implies that
$\mc{I}_{\tau}(p)\subseteq \mc{I}_\tau(q)$, where the set
$\mc{I}_\tau(p)$ of \emph{weak} initial actions is $\mc{I}_\tau(p)
= \{\alpha \in A_{\tau} \mid p\Rightarrow\mv{\alpha}\}$} that
contains the strong ready simulation preorder $\sqsubseteq_{\rm RS}$
and satisfies T2 (the second $\tau$-law of CCS: $\tau x \approx
\tau x + x$), and let $E$ be a sound and ground-complete
axiomatization of $\sqsubseteq$. Then $\mc{A}(E)$ is a sound and
ground-complete axiomatization of the kernel of $\sqsubseteq$.
Moreover, if $E$ is $\omega$-complete, then so is $\mc{A}(E)$.
\end{theorem}
It is straightforward to check that weak failures meets the
prerequisites of Theo.~\ref{alg}, and thus we can run the
algorithm  and obtain the axiomatization in Tab.~\ref{tab-aux} for
weak failures equivalence.
\begin{table}
\vspace{-5pt}
\begin{center}
 \begin{tabular}{l@{\qquad}rcl}
 WF1   & $ax+ay$  & $\approx$ & $a(\tau x+\tau y)$\\
 WF2$^a$ & $\tau(x+y)+\tau x +y$  & $\approx$ & $\tau x+ y$\\
 WF2$^b$ & $\alpha(\tau(x+y)+z) + \alpha (\tau x+y +z) $  & $\approx$ & $\alpha(\tau x+ y+z)$\\
 WF3$^a$ & $x+\tau x+y$  & $\approx $ & $ \tau x+  y$\\
 WF3$^b$ & $\alpha(x+z)+\alpha(\tau x+y+z)$  & $\approx $ & $ \alpha(\tau x+  y+z)$\\
 RS   & $\beta(\alpha x + z) + \beta(\alpha x + \alpha y + z)$  & $\approx $ &
 $\beta(\alpha x + \alpha y + z)$\\
 WF$_A^{~~a}$  & $\sum_{a\in A} ax_a + \sum_{a\in A} ax_a + y$  & $\approx $ &
 $\sum_{a\in A} ax_a + y$\\
 WF$_A^{~~b}$  & $\beta(\sum_{a\in A} ax_a + z) + \beta(\sum_{a\in A} ax_a + y+z)$  & $\approx $ & $\beta(\sum_{a\in A} ax_a + y+z)$
\end{tabular}
\end{center}\caption{Axiomatization generated from the algorithm}\label{tab-aux}
\vspace{-5mm}
\end{table}
\noindent After simplification and omission of redundant axioms,
we obtain the axiomatization in Tab.~\ref{tab5}.
\begin{table}
\vspace{-5pt}
\begin{center}
 \begin{tabular}{l@{\qquad}rcl}
 WF1  & $ax+ay$  & $\approx$ & $a(\tau x+\tau y)$\\
 WFE2 & $\tau(x+y)+\tau x$  & $\approx$ & $\tau x+ y$\\
 WFE3 & $ax+\tau(ay+z)$  & $\approx$ & $\tau(a x+ a y+z)$\\
 WFE$_A$ & $\tau(\sum_{a\in A} ax_a + z) + \tau(\sum_{a\in A} ax_a + y+z)$
  & $\approx $ & $\tau(\sum_{a\in A} ax_a + y+z)$
\end{tabular}
\end{center}
\caption{Axiomatization for weak failures equivalence} \label{tab5}
\vspace{-5mm}
\end{table}

  \begin{lemma} \label{lem:four}
     The axioms in \emph{Tab.~\ref{tab-aux}} are
     derivable from the axioms in
     \emph{Tab.~\ref{tab5}} together with A1-4.
  \end{lemma}
  \begin{proof}
%
   WF1 is unmodified. WF2$^a$ and WF3$^a$ can be trivially
  derived from WFE2. WF$_A^{~~a}$ is derivable using A3.

  To proceed, we have that WFE2$\vdash \tau\tau x \approx \tau x$ (namely by
  substituting $\tau x$ for $y$ and invoking D1) and hence also WFE2$\vdash$D2
  (namely by substituting $\tau x $ for $x$ in WFE2 and invoking D1);
  using D2, the instances of WF2$^b$ and WF3$^b$ with $\alpha=\tau$, as well as the
  instance of RS with $\beta=\alpha=\tau$,
  are derivable from WFE2.

  The instances of WF2$^b$ and WF3$^b$ with $\alpha\neq\tau$,
  are derivable from WF1 and the instances with $\alpha=\tau$;
  the same holds for the instances of RS and WF$_A^{~~b}$ with
  $\beta\neq\tau$.

  Finally in the remaining instances of RS
  (with $\beta=\tau$ and $\alpha=a\in A$), we have WFE2$\vdash \tau(ax+z)+\tau(ax+ay+z)\approx
  \tau(ax+z)+ay$,
  and thus it can be derived from WFE3.  The instance of
  WF$_A^{~~b}$ with $\beta=\tau$ is exactly WFE$_A$. \qed

%
  \end{proof}

The axioms WF1, \mbox{WFE2-3} already appeared in \cite{Gla97}.
A1-4+WF1+WFE2-3 is sound and ground-complete for BCCS modulo
$\equiv_{\rm WF}$ (see also \cite{Gla97,CFG08b}). By
Theo.~\ref{theo:A-infinite} and Theo.~\ref{theo:A-finite}
(together with Lem.~\ref{lem:four}), we have:

\begin{corollary}
  If $|A|=\infty$, then the axiomatization \mbox{\rm A1-4+WF1+WFE2-3} is
  $\omega$-complete for $\mathrm{BCCS}(A)$ modulo $\equiv_{\rm WF}$.
\end{corollary}

\begin{corollary}
  If $|A| <\infty$, then the axiomatization \mbox{\rm A1-4+WF1+WFE2-3+WFE$_A$} is
  $\omega$-complete for
   $\mathrm{BCCS}(A)$ modulo $\equiv_{\rm WF}$.
\end{corollary}

\section{Weak Impossible Futures Semantics} \label{sec6}

\emph{Weak impossible futures} semantics is closely related to
weak failures semantics. Only, instead of the set of actions in
the second argument of a weak failure pair (see
Def.~\ref{def:weak-failures}), an impossible future pair contains
a set of \emph{traces}.

\begin{definition}[Weak impossible futures]\rm\label{wif}
\vspace{-1ex}
\begin{itemize}
  \item A sequence $a_1\cdots a_k \in A^*$, with $k\geq 0$, is a
  \emph{trace} of a process $p_0$ if there is a path
  $p_0\Rightarrow \mv{a_1} \Rightarrow \cdots \Rightarrow \mv{a_k} \Rightarrow p_k$;
  it is a \emph{completed trace} of $p_0$ if moreover $\mathcal{I}(p_k)=\emptyset$.
  Let $\mc{T}(p)$ denote the set of traces of process $p$, and
  $\mc{CT}(p)$ its set of completed traces.

  \item A pair $(a_1\cdots a_k,B)$, with $k\geq 0$ and $B \subseteq A^*$,
  is a \emph{weak impossible future} of a process $p_0$ if there is a
  path $p_0\Rightarrow \mv{a_1} \Rightarrow \cdots \Rightarrow \mv{a_k} \Rightarrow p_k$
  with $\mc{T}(p_k) \cap B = \emptyset$.

  \item The \emph{weak impossible futures preorder} $\sqsubseteq_{\rm WIF}$ is
  given by $p\sqsubseteq_{\rm WIF} q$ iff (1) the weak impossible
  futures of $p$ are also weak impossible futures of $q$,
  (2) $\mc{T}(p)=\mc{T}(q)$ and
  (3) $p\mv{\tau}$ implies that $q\mv{\tau}$.

  \item \emph{Weak impossible futures equivalence} $\equiv_{\rm WIF}$ is defined
  as $\sqsubseteq_{\rm WIF} \cap \sqsubseteq_{\rm WIF}^{-1}$.
\end{itemize}
\end{definition}
$\sqsubseteq_{\rm WIF}$ is a precongruence, and $\equiv_{\rm WF}$ a
congruence, for BCCS \cite{VM01}.
   The requirement (2) $\mc{T}(p)=\mc{T}(q)$ is necessary for this
   precongruence property. Without it we would have $\tau a\nil \sqsubseteq
   \tau a\nil + b\nil$ but $c(\tau a\nil) \not\sqsubseteq c(\tau a\nil + b\nil)$.

A sound and ground-complete axiomatization for $\sqsubseteq_{\rm WIF}$
is obtained by replacing axiom WF3 in Tab.~\ref{tab1} by the
following axiom (cf.\ \cite{VM01}, where a slightly more
complicated, but equivalent, axiomatization is given):
\[
 \mbox{WIF3}~~~~  x  ~\preccurlyeq~ \tau x
\]
However, surprisingly, there is no finite sound and
ground-complete axiomatization for $\equiv_{\rm WIF}$.
We will show this in Sec.~\ref{nonax}. A similar
difference between the impossible futures preorder and equivalence
in the concrete case (so in the absence of $\tau$) was found
earlier in \cite{CF08}. We note that, since weak impossible
futures semantics is not coarser than ready simulation semantics,
the algorithm from \cite{AFI07,FG07,CFG08b} to generate an
axiomatization for the equivalence from the one for the preorder,
does not work in this case.

In Sec.~\ref{sec:42} we establish that the sound and ground-complete
axiomatization for BCCS modulo $\sqsubseteq_{\rm WIF}$ is
$\omega$-complete in case $|A|=\infty$, and in Sec.~\ref{sec:43} that
there is no such finite basis for the inequational theory of BCCS
modulo $\sqsubseteq_{\rm WIF}$ in case $|A|<\infty$. Again, these
results correspond to (in)axiomatizability results for the impossible
futures preorder in the concrete case \cite{CF08}, with very similar
proofs.

\subsection{Nonexistence of an Axiomatization for Equivalence}\label{nonax}
We now prove that for any (nonempty) $A$ there does \emph{not} exist
any finite, sound, ground-complete axiomatization for
$\mathrm{BCCS}(A)$ modulo $\equiv_{\rm WIF}$. The cornerstone for this
negative result is the following infinite family of closed equations,
for $m\geq 0$:
\[ \tau a^{2m}\nil+\tau(a^m\nil+a^{2m}\nil) ~\approx~ \tau(a^m\nil+a^{2m}\nil) \]
It is not hard to see that they are sound modulo $\equiv_{\rm WIF}$.
We start with a few lemmas.

\begin{lemma}\label{res}
If $p\sqsubseteq_{\rm WIF} q$ then $\mc{CT}(p) \subseteq \mc{CT}(q)$.
\end{lemma}

\begin{proof}
A process $p$ has a completed trace $a_1 \cdots a_k$ iff it has a weak
impossible future $(a_1 \cdots a_k,A)$.
\qed
\end{proof}

\begin{lemma} \label{newvariable}
Suppose $t\sqsubseteq_{\rm WIF} u$. Then for any $t'$ with
$t\Rightarrow\mv{\tau} t'$ there is some $u'$ with
$u\Rightarrow\mv{\tau} u'$ such that $\var(u')\subseteq \var(t')$.
\end{lemma}

\begin{proof}
Let $t\Rightarrow\mv{\tau} t'$. Fix some $m>|t|$, and consider
the closed substitution $\rho$ defined by $\rho(x)=\nil$ if
$x\in\var(t')$ and $\rho(x)=a^m\nil$ if $x\not\in\var(t')$. Since
$\rho(t)\Rightarrow\rho(t')$ with $|\rho(t')|=|t'|<m$, and
$\rho(t)\sqsubseteq_{\rm WIF}\rho(u)$, clearly $\rho(u)\Rightarrow q$
for some $q$ with $|q|<m$. From the definition of $\rho$ it then
follows that there must exist $u\Rightarrow u'$ with
$\var(u')\subseteq \var(t')$. In case $u \Rightarrow\mv{\tau} u'$ we
are done, so assume $u'=u$. Let $\sigma$ be the substitution with
$\sigma(x)=\nil$ for all $x\in V$. Since $\sigma(t) \mv{\tau}$ and $t
\sqsubseteq_{\rm WIF} u$ we have $\sigma(u)\mv{\tau}$, so $u \mv{\tau}
u''$ for some $u''$. Now $\var(u'') \subseteq \var(u) = \var(u')
\subseteq \var(t')$.\qed
\end{proof}
\pagebreak[3]

\begin{lemma} \label{key2}
Assume that, for terms $t,u$, closed substitution $\sigma$, action $a$
and integer $m$:

\begin{enumerate}\vspace{-1ex}
  \item $t \equiv_{\rm WIF} u$;

  \item $m > |u|$;

  \item $\mc{CT}(\sigma(u)) \subseteq \{a^{m}, a^{2m}\}$; and

  \item there is a closed term $p'$ such that
  $\sigma(t)\Rightarrow\mv{\tau} p'$ and $\mc{CT}(p') = \{a^{2m}\}$.
\vspace{-1ex}
\end{enumerate}
Then there is a closed term $q'$ such that $\sigma(u)
\Rightarrow\mv{\tau} q'$ and $\mc{CT}(q') = \{a^{2m}\}$.
\end{lemma}

\begin{proof}
According to proviso (4) of the lemma, we can distinguish two
cases.

\begin{itemize}
  \item There exists some $x\in V$ such that $t \Rightarrow t'$ with
  $t'=t''+x$ and $\sigma(x)\Rightarrow\mv{\tau} p'$ where $\mc{CT}(p')=\{a^{2m}\}$.
  Consider the closed substitution $\rho$ defined by $\rho(x)=a^m\nil$
  and $\rho(y)=\nil$ for any $y\neq x$. Then $a^m \in \mc{CT}(\rho(t))
  = \mc{CT}(\rho(u))$, using Lem.~\ref{res}, and this is only
  possible if $u \Rightarrow u'$ for some $u'=u''+x$. Hence
  $\sigma(u)\Rightarrow\mv{\tau} p'$.

  \medskip

  \item $t\Rightarrow\mv{\tau} t'$ with $\mc{CT}(\sigma(t'))=\{a^{2m}\}$.
  Since $|t'|\leq|t|=|u|<m$, clearly, for any \mbox{$x\mathbin\in \var(t')$},
  either $|\sigma(x)|\mathbin=0$ or $\norm(\sigma(x))\mathbin>m$,
  where $\norm(p)$ denotes the length of the shortest completed trace of $p$.
  Since $t\equiv_{\rm WIF} u$, by Lem.~\ref{newvariable},
  $u\mathbin\Rightarrow\mv{u} u'$ with $\var(u')\subseteq \var(t')$. Hence,
  for any $x\mathbin\in \var(u')$, either $|\sigma(x)|\mathbin=0$
  or $\norm(\sigma(x))\mathbin>m$.  Since $|u'|\mathbin<m$,
  $a^m\mathbin{\notin} \mc{CT}(\sigma(u'))$.  It follows from
  $\mc{CT}(\sigma(u))\subseteq\{a^{m}, a^{2m}\}$ that
  $\mc{CT}(\sigma(u'))= \{a^{2m}\}$. And $u\Rightarrow\mv{\tau} u'$
  implies $\sigma(u)\Rightarrow\mv{\tau} \sigma(u')$.\hfill$\Box$
\end{itemize}
\end{proof}

\begin{lemma} \label{key2cont}
Assume that, for $E$ an axiomatization sound for
$\sqsubseteq_{\rm WIF}$, closed terms $p,q$, closed
substitution $\sigma$, action $a$ and integer $m$:
\begin{enumerate}\vspace{-1ex}
  \item $E\vdash  p \approx q$;

  \item $m > \max \{ |u|\mid t \approx u \in  E\}$;

  \item $\mc{CT}(q) \subseteq \{a^{m},a^{2m}\}$; and

  \item there is a closed term $p'$ such that $p\Rightarrow\mv{\tau} p'$ and
  $\mc{CT}(p') = \{a^{2m}\}$.
\vspace{-1ex}
\end{enumerate}
Then there is a closed term $q'$ such that $q \Rightarrow\mv{\tau} q'$ and
$\mc{CT}(q') = \{a^{2m}\}$.
\end{lemma}

\begin{proof}

By induction on the derivation of $E\vdash  p \approx q$.

\begin{itemize}

\item Suppose $E\vdash p\approx q$ because $\sigma(t)=p$ and
$\sigma(u)=q$ for some $t\approx u\in E$ or $u\approx t\in E$ and
closed substitution $\sigma$. The claim then follows by Lem.~\ref{key2}.

\medskip

\item Suppose $E\vdash p\approx q$ because $E\vdash p\approx r$ and
$E\vdash r\approx q$ for some $r$. Since $r\equiv_{\rm WIF} q$, by
proviso (3) of the lemma and Lem.~\ref{res},
$\mc{CT}(r)\subseteq \{a^{m},a^{2m}\}$. Since there is a $p'$
such that $p\Rightarrow\mv{\tau} p'$ with $\mc{CT}(p') = \{a^{2m}\}$,  by
induction, there is an $r'$ such that $r\Rightarrow\mv{\tau} r'$ and
$\mc{CT}(r') = \{a^{2m}\}$. Hence, again by induction, there is a
$q'$ such that $q\Rightarrow\mv{\tau} q'$ and $\mc{CT}(q') = \{a^{2m}\}$.

\medskip

\item Suppose $E\vdash p\approx q$ because $p=p_1+p_2$ and
$q=q_1+q_2$ with $E\vdash p_1\approx q_1$ and $E\vdash p_2\approx
q_2$. Since there is a $p'$ such that $p\Rightarrow\mv{\tau} p'$ and
$\mc{CT}(p') = \{a^{2m}\}$, either $p_1\Rightarrow \mv{\tau} p'$ or
$p_2\Rightarrow\mv{\tau} p'$. Assume, without loss of generality, that
$p_1\Rightarrow\mv{\tau} p'$. By induction, there is a $q'$ such that
$q_1\Rightarrow\mv{\tau} q'$ and $\mc{CT}(q') = \{a^{2m}\}$.
Now $q\Rightarrow\mv{\tau} q'$.

\medskip

\item Suppose $E\vdash p \approx q$ because $p=cp_1$ and $q=cq_1$
with $c\in A$ and $E\vdash p_1\approx q_1$. In this case, proviso (4)
of the lemma can not be met.

\medskip

\item Suppose $E\vdash p \approx q$ because $p=\tau p_1$ and $q=\tau
q_1$ with $E\vdash p_1\approx q_1$. By proviso (4) of the lemma,
either $\mc{CT}(p_1)=\{a^{2m}\}$ or there is a $p'$ such that
$p_1\Rightarrow\mv{\tau} p'$ and $\mc{CT}(p') = \{a^{2m}\}$.
In the first case, $q \Rightarrow\mv{\tau} q_1$ and $\mc{CT}(q_1) = \{a^{2m}\}$
by Lem.~\ref{res}. In the second, by induction, there is a $q'$
such that $q_1\Rightarrow\mv{\tau} q'$ and $\mc{CT}(q') = \{a^{2m}\}$.
Again $q\Rightarrow\mv{\tau} q'$.
\qed
\end{itemize}
\end{proof}

\begin{theorem}
There is no finite, sound, ground-complete axiomatization for
$\mathrm{BCCS}(A)$ modulo $\equiv_{\rm WIF}$.
\end{theorem}

\begin{proof}
Let $E$ be a finite axiomatization over $\mathrm{BCCS}(A)$ that is
sound modulo $\equiv_{\rm WIF}$. Let $m$ be greater than the depth
of any term in $E$. Clearly, there is no term $r$ such that
$\tau(a^m\nil+a^{2m}\nil) \Rightarrow\mv{\tau} r$ and
$\mc{CT}(r)=\{a^{2m}\}$. So according to Lem.~\ref{key2cont}, the
closed equation $\tau a^{2m}\nil+\tau(a^m\nil+a^{2m}\nil) \approx
\tau(a^m\nil+a^{2m}\nil)$ cannot be derived from $E$. Nevertheless, it is
valid modulo $\equiv_{\rm WIF}$. \qed
\end{proof}

In the same way as above, one can establish the nonderivability of
the equations $a^{2m+1}\nil+a(a^m\nil+a^{2m}\nil) ~\approx~
a(a^m\nil+a^{2m}\nil)$ from any given finite equational
axiomatization sound for $\equiv_{\rm WIF}$. As these equations
are valid modulo (strong) 2-nested simulation equivalence, this
negative result applies to all BCCS-congruences that are at least
as fine as weak impossible futures equivalence and at least as
coarse as strong 2-nested simulation equivalence. Note that
the corresponding result of \cite{AFGI04} can be inferred.

\subsection{A Finite Basis for Preorder if $|A|=\infty$} \label{sec:42}


In this section, we show that A1-4+WF1-2+WIF3 is $\omega$-complete
in case $|A|=\infty$. Note that this result was originally
obtained in \cite{VM01}. However, our proof is much simpler.
First, let us note that A1-4+WF1-2+WIF3 $\vdash$ D1, D2, D5.
\begin{lemma}
  For any closed terms $p, q$, if $p\sqsubseteq_{\rm WIF} q$, then
  \mbox{\rm A1-4+WF1-2+WIF3} $\vdash p\preccurlyeq q$.
\end{lemma}

\begin{proof} Let $p\sqsubseteq_{\rm WIF} q$. We prove $\vdash
  p\preccurlyeq q$ by induction on $|p|+|q|$.  We distinguish two
  cases:
  \begin{itemize}
     \item $q\not\!\mv{\tau}$. Then $p\not\!\mv{\tau}$ since $p\sqsubseteq_{\rm WIF} q$. Suppose
      $p=\sum_{i\in I}a_ip_i$ and $q=\sum_{j\in J}b_j q_j$. Clearly,
      we have $\mc{I}(p)=\mc{I}(q)$.
    By D5, we have
    \[ \vdash p\approx \sum_{a\in \mc{I}(p)} a(\sum_{a_i=a, i\in I}\tau p_i)\]
    and
    \[ \vdash q\approx \sum_{a\in \mc{I}(p)} a(\sum_{b_j=a, j\in J}\tau q_j)\]
    Since $p\sqsubseteq_{\rm WIF} q$, for each $a\in \mc{I}(p)$, the following relation holds:
    \[  \sum_{a_i=a, i\in I}\tau p_i \sqsubseteq \sum_{b_j=a, j\in J} \tau q_j\]
    By induction,
    \[ \vdash \sum_{a_i=a, i\in I}\tau p_i \preccurlyeq \sum_{b_j=a, j\in J}\tau q_j\]
    and thus
    \[ \vdash a(\sum_{a_i=a, i\in I}\tau p_i) \preccurlyeq a(\sum_{b_j=a, j\in J}\tau q_j)\]
    Summing these up for $a\in \mc{I}(p)$, we obtain that
    \[ \vdash p\preccurlyeq q\]

\item $q\mv{\tau}$. By D2, we can write $ p\approx
\sum_{i\in I}\alpha_ip_i$ and $q\approx \sum_{j\in J} \beta_jq_j$ such that
for each $\alpha_i=\tau$ (resp. $\beta_j=\tau$), $p_i\not\!\mv{\tau}$
(resp. $q_j\not\!\mv{\tau}$).
Applying D1, for each $i\in I$ with $\alpha_i=\tau$, the summands of
$p_i$ are also made summands of $p$, and likewise for $q$.\hfill (*)

\medskip

For each $i\in I$ with $\alpha_i=\tau$ we have $p\mv{\tau} p_i$. Since
$p\sqsubseteq_{\rm WIF}q$ and no $q_j$ with $\beta_j=\tau$ contains a
$\tau$-summand, either $\mc{T}(q)\subseteq\mc{T}(p_i)$ or there exists
\raisebox{0pt}[0pt][0pt]{$q\mv{\tau}q_j$} such that $\mc{T}(q_j)\subseteq \mc{T}(p_i)$.
Since \raisebox{0pt}[0pt][0pt]{$q\mv{\tau}$}, in either case there exists some $j_i\in J$ such
that $b_{j_i}=\tau$ and  $\mc{T}(q_{j_i})\subseteq \mc{T}(p_i)$. It follows that
      \[ p_i \sqsubseteq_{\rm WIF} p_i + q_{j_i}\]
Since $p_i \not\!\mv{\tau}$ and $q_{j_i} \not\!\mv{\tau}$, by the previous
case,
      \[ \vdash p_i \preccurlyeq  p_i + q_{j_i}\]
Hence by WF2,
\[ \vdash \tau p_i \preccurlyeq  \tau (p_i + q_{j_i}) \preccurlyeq p_i
+\tau q_{j_i} \]
and thus
\[ \vdash p= \sum_{\alpha_i=\tau} \tau p_i + \sum_{a\in \mc{I}(p)}\sum_{\alpha_i=a, i\in I} ap_i
\preccurlyeq  \sum_{\alpha_i=\tau} (p_i +\tau q_{j_i}) + \sum_{a\in
\mc{I}(p)}\sum_{\alpha_i=a, i\in I} ap_i\]
By (*),
\[\vdash \sum_{\alpha_i=\tau} p_i + \sum_{a\in
\mc{I}(p)}\sum_{\alpha_i=a, i\in I} ap_i \approx \sum_{a\in
\mc{I}(p)}\sum_{\alpha_i=a, i\in I} ap_i \]
Since $p\sqsubseteq_{\rm WIF} q$, $\mc{I}(p)=\mc{I}(q)$.
Using (*), it is easy to see that for each $a\in \mc{I}(p)$,
\[ \sum_{\alpha_i=a, i\in I} ap_i \sqsubseteq_{\rm WIF}  \sum_{\beta_j=a,j\in J} aq_j\]
So by the previous case,
\[ \vdash \sum_{\alpha_i=a, i\in I} ap_i \preccurlyeq  \sum_{\beta_j=a,j\in J} aq_j\]
It follows that
\[ \vdash p \preccurlyeq \sum_{\alpha_i=\tau} \tau q_{j_i} + \sum_{a\in
\mc{I}(p)}\sum_{\alpha_i=a, i\in I} ap_i \preccurlyeq \sum_{\alpha_i=\tau}
\tau q_j + \sum_{a\in \mc{I}(p)}\sum_{\beta_j=a,j\in J} aq_j\]
By WIF3,
\[\vdash \sum_{\alpha_i=\tau} \tau q_j + \sum_{a\in
\mc{I}(p)}\sum_{\beta_j=a,j\in J} aq_j \preccurlyeq q \]
Hence \hfill $\vdash p \preccurlyeq q$\hfill\mbox{} \qed
  \end{itemize}
\end{proof}
With this ground-completeness result at hand, it is
straightforward to apply the \emph{inverted substitution}
technique of Groote \cite{Gro90} to derive (see also \cite{CF08}):
\begin{theorem}
If $|A|=\infty$, then \mbox{\rm A1-4+WF1-2+WIF3} is $\omega$-complete
for $\mathrm{BCCS}(A)$ modulo $\sqsubseteq_{\rm WIF}$.
\end{theorem}
\begin{proof}
Given an inequational axiomatization $E$ and open terms $t,u$ such
that $E \vdash \sigma(t) \preccurlyeq \sigma(u)$ for all closed
substitutions $\sigma$, the technique of inverted substitutions is a
method to prove $E \vdash t \preccurlyeq u$. It does so by means of a
closed substitution $\rho$ encoding open terms into closed terms, and
an decoding operation $R$ that turns closed terms back into
open terms. By assumption we have $E \vdash \rho(t) \preccurlyeq
\rho(u)$. The pair $(\rho,R)$ should be chosen in such a way that, in
essence, applying $R$ to all terms occurring in a proof of $\rho(t)
\preccurlyeq \rho(u)$ yields a proof of $t \preccurlyeq u$.  As
observed in \cite{Gro90}, this technique is applicable when three
conditions are met, one of which being that $R(\rho(t))=t$ and
$R(\rho(u))=u$. In fact, \cite{Gro90} dealt with equational logic
only, but the very same reasoning applies to inequational logic.

Here we use the same pair $(\rho,R)$ that was used by Groote to obtain
most of the applications of the technique in \cite{Gro90}---it could
be called the \emph{default} (inverted) substitution. It is obtained
by selecting for each variable $x\in V$ an action $a_x\in A$, not
occurring in $t$ or $u$. This is possible because $|A|=\infty$. Now
the default substitution $\rho$ is given by $\rho(x)=a_x\nil$ and the
default inverted substitution $R$ replaces any maximal subterm of the
form $a_x p$ into the variable $x$. Groote showed that with this
particular (inverted) substitution, 2 out of his 3 conditions are
always met, and the third one simply says that for each axiom $t
\preccurlyeq u$ in $E$ we should have that $E \vdash R(t) \preccurlyeq
R(u)$.  This condition is clearly met for the axioms A1-4+WF1-2+WIF3,
and hence this axiomatization is $\omega$-complete.\qed
\end{proof}
Note that we could have used the same method to obtain
Theo.~\ref{theo:A-infinite}, but not Theo.~\ref{theo:A-finite}.

\subsection{Nonexistence of a Finite Basis for Preorder if
$|A|<\infty$} \label{sec:43}

\subsubsection{$1< |A| <\infty$.}
We prove that the inequational theory of $\mathrm{BCCS}(A)$ modulo
$\sqsubseteq_{\rm WIF}$ does \emph{not} have a finite basis in case of
a finite alphabet with at least two elements.
The cornerstone for this negative result is the following
infinite family of inequations, for $m\geq 0$:
\[\tau(a^mx) +  \Phi_m ~\preccurlyeq~ \Phi_m \]
with
\[\Phi_m ~=~ \tau(a^mx+x) + \sum_{b\in A} \tau(a^mx+a^mb\nil)\]
It is not hard to see that these inequations are sound modulo
$\sqsubseteq_{\rm WIF}$. Namely, given a closed substitution $\rho$,
we have $\mc{T}(\rho(\tau(a^mx))) \subseteq \mc{T}(\rho(\Phi_m))$ and
$\rho(\Phi_m)\mv{\tau}$.
To argue that $\rho(\tau(a^mx) + \Phi_m)$ and $\rho(\Phi_m)$ have the
same impossible futures, we only need to consider the
transition $\rho(\tau(a^mx)+\Phi_m)\mv{\tau}a^m\rho(x)$ (all other
cases being trivial). If
$\rho(x)=\nil$, then $\rho(\Phi_m)\mv{\tau}a^m\nil + \nil$ generates
the same impossible futures $(\varepsilon,B)$.  If, on the other hand,
$b\mathbin\in\mc{I}(\rho(x))$ for some $b\mathbin\in A$, then
$\mc{T}(a^m\rho(x)+a^mb\nil)=\mc{T}(a^m\rho(x))$, so
$\rho(\Phi_m)\mv{\tau} a^m\rho(x)+a^mb\nil$ generates the same
impossible futures $(\varepsilon,B)$.

We have already defined the traces and completed traces of closed
terms. Now we extend these definitions to open terms by allowing
(completed) traces of the form $a_1 \cdots a_k x \in A^\ast V$.
We do this by treating each variable occurrence $x$ in a term as if it
were a subterm $x\nil$ with $x$ a visible action, and then apply
Def.~\ref{wif}. Under this convention,
$\mc{CT}(\Phi_m) =\{a^{m}x, x, a^{m}b \mid b\mathbin\in A\}$.
We write $\mc{T}_V(t)$ for the set of traces of $t$ that end in a
variable, and $\mc{T}_A(t)$ for ones that end in an action.

\begin{observation}\label{traces-substitution}
Let $m>|t|$ or $a_m \in V$.
Then $a_1\cdots a_m \in \mc{T}(\sigma(t))$ iff there is a $k < m$
and $y\mathbin\in V$ such that $a_1\cdots a_k y \in \mc{T}_V(t)$ and
$a_{k+1} \cdots a_m \in \mc{T}(\sigma(y))$.
\end{observation}

\begin{lemma}\label{trace-preservation}
If $|A|>1$ and $t \sqsubseteq_{\rm WIF} u$ then
$\mc{T}_A(t)= \mc{T}_A(u)$ and $\mc{T}_V(t)= \mc{T}_V(u)$.
\end{lemma}

\begin{proof}
Let $\sigma$ be the closed substitution defined by
$\sigma(x)\mathbin=\nil$ for all $x\mathbin\in V$.  Then
$t\sqsubseteq_{\rm WIF} u$ implies $\sigma(t)\sqsubseteq_{\rm WIF}
\sigma(u)$ and hence $\mc{T}_A(t)=\mc{T}(\sigma(t))=
\mc{T}(\sigma(u))=\mc{T}_A(u)$ by Def.~\ref{wif}.

For the second statement fix distinct actions $a,b\mathbin\in A$ and
an injection $\ulcorner \cdot \urcorner: V \rightarrow
\mathbb{Z}_{>0}$ (which exists because $V$ is countable). Let
$m=|u|+1=|t|+1$. Define the closed substitution $\rho$ by
$\rho(z)=a^{\ulcorner\! z\!\urcorner{\cdot}m}b\nil$ for all $z\in V$.
Again, by Def.~\ref{wif}, $t\sqsubseteq_{\rm WIF} u$ implies
$\mc{T}(\rho(t))= \mc{T}(\rho(u))$. By Obs.~\ref{traces-substitution},
for all terms $v$ we have $a_1\cdots a_k y \in \mc{T}_V(v)$ iff $a_1\cdots a_k
a^{\ulcorner\!  y\!\urcorner{\cdot}m}b \in \mc{T}(\rho(v))$ with
$k\mathbin<m$.  Hence $\mc{T}_V(v)$ is completely determined by
$\mc{T}(\rho(v))$ and thus $\mc{T}_V(t) \mathbin= \mc{T}_V(u)$.\qed
\end{proof}

\begin{lemma} \label{newvariable3}
Let $|A|>1$. Suppose $t\sqsubseteq_{\rm WIF} u$ and $t\Rightarrow\mv{\tau} t'$.
Then there is a term $u'$ such that $u\Rightarrow\mv{\tau} u'$ and
$\mc{T}_V(u')\subseteq \mc{T}_V(t')$.
\end{lemma}

\begin{proof}
Define $\rho$ exactly as in the previous proof. Since
$\rho(t)\Rightarrow\rho(t')$ and $t\sqsubseteq_{\rm WIF} u$ there must
be a $u'$ with $\rho(u) \Rightarrow q$ and $\mc{T}(q)\subseteq
\mc{T}(\rho(t'))$. Since $\rho(x)$ is $\tau$-free for $x\in V$ it must
be that $q=\rho(u')$ for some term $u'$ with $u \Rightarrow u'$.
Given the relationship between $\mc{T}_V(v)$ and $\mc{T}(\rho(v))$ for
terms $v$ observed in the previous proof, it follows that
$\mc{T}_V(u')\subseteq\mc{T}_V(t')$. In case $u \Rightarrow\mv{\tau}
u'$ we are done, so assume $u'=u$. Let $\sigma$ be the substitution
with $\sigma(x)=\nil$ for all $x\in V$. Since $\sigma(t) \mv{\tau}$
and $t \sqsubseteq_{\rm WIF} u$ we have $\sigma(u)\mv{\tau}$, so $u
\mv{\tau} u''$ for some $u''$. Now $\mc{T}_V(u'') \subseteq \mc{T}_V(u) =
\mc{T}_V(u') \subseteq \mc{T}_V(t')$.
\qed
\end{proof}

\begin{lemma} \label{no-traces1}
  Let $|A|>1$. Assume that, for some terms $t,u$, substitution
  $\sigma$, action $a$ and integer $m$:
  \begin{enumerate}\vspace{-1ex}
  \item $t \sqsubseteq_{\rm WIF} u$;

  \item $m \geq |u|$; and

  \item $\sigma(t)\Rightarrow\mv{\tau} \hat t$ for a term $\hat t$ without
  traces $ax$ for $x\mathbin\in V$ or $a^mb$ for $b\mathbin\in A$.
\vspace{-1ex}\end{enumerate}
Then $\sigma(u)\Rightarrow\mv{\tau} \hat u$ for a term $\hat u$ without
  traces $ax$ for $x\mathbin\in V$ or $a^mb$ for $b\mathbin\in A$.
\end{lemma}
\begin{proof}
Based on proviso (3) there are two cases to consider.\vspace{-1ex}
\begin{itemize}
\item $y \in \mc{T}_V(t)$ for some $y\mathbin\in V$ and
  $\sigma(y)\Rightarrow\mv{\tau} \hat t$.
  In that case $y \mathbin\in \mc{T}_V(u)$ by
  Lem.~\ref{trace-preservation}, so $\sigma(u)\Rightarrow\mv{\tau} \hat t$.
\item $t \Rightarrow\mv{\tau} t'$ for some term $t'$ such that $\hat t
  = \sigma(t)$.  By Lem.~\ref{newvariable3} there is a term $u'$ with
  $u\Rightarrow\mv{\tau} u'$ and $\mc{T}_V(u')\subseteq \mc{T}_V(t')$.
  Take $\hat u = \sigma(u')$. Clearly $\sigma(u)\Rightarrow\mv{\tau}
  \sigma(u')$.  Suppose $\sigma(u')$ would have a trace $a^mb$. Then,
  by Obs.~\ref{traces-substitution}, there is a $k \leq m$ and
  $y\mathbin\in V$ such that $a^k y \in \mc{T}_V(u')$ and
  $a^{m-k}b \in \mc{T}(\sigma(y))$. Since $\mc{T}_V(u')\subseteq
  \mc{T}_V(t')$ we have $a^mb \in \mc{T}(\sigma(t'))$, which is a
  contradiction. The case $ax \in \mc{T}(\sigma(u))$ is dealt with in
  the same way.\qed
\end{itemize}
\end{proof}

\begin{lemma} \label{no-traces2}
  Let $|A|>1$ and let $E$ be an axiomatization sound for
  $\sqsubseteq_{\rm WIF}$. Assume that, for some terms $v,w$, action
  $a$ and integer $m$:
  \begin{enumerate}\vspace{-1ex}
  \item $E\vdash  v \preccurlyeq w$;

  \item $m \geq \max \{|u| \mid t \preccurlyeq u \in  E\}$; and

  \item $v\Rightarrow\mv{\tau} \hat v$ for a term $\hat v$ without
  traces $ax$ for $x\mathbin\in V$ or $a^mb$ for $b\mathbin\in A$.
\vspace{-1ex}\end{enumerate}
Then $w\Rightarrow\mv{\tau}  \hat w$ for a term $\hat w$ without
  traces $ax$ for $x\mathbin\in V$ or $a^mb$ for $b\mathbin\in A$.
\end{lemma}

\begin{proof}
By induction on the derivation of $E\vdash v \preccurlyeq w$.
\begin{itemize}

\item Suppose $E\vdash v\preccurlyeq w$ because $\sigma(t)=v$ and
$\sigma(u)=w$ for some $t\preccurlyeq u\in E$ and substitution
$\sigma$. The claim then follows by Lem.~\ref{no-traces1}.

\medskip

\item Suppose $E\vdash v\preccurlyeq w$ because $E\vdash v\preccurlyeq u$
and $E\vdash u\preccurlyeq w$ for some $u$. By induction, $u
\Rightarrow\mv{\tau} \hat u$ for a term $\hat u$ without traces $ax$ or
$a^mb$. Hence, again by induction, \mbox{$w\Rightarrow\mv{\tau} \hat w$}
for a term $\hat w$ without traces $ax$ or $a^mb$.

\medskip

\item Suppose $E\vdash v\preccurlyeq w$ because $v=v_1+v_2$ and
$w=w_1+w_2$ with $E\vdash v_1 \preccurlyeq w_1$ and $E\vdash
v_2\preccurlyeq w_2$. Since $v \Rightarrow\mv{\tau} \hat v$, either
$v_1 \Rightarrow\mv{\tau} \hat v$ or $v_2 \Rightarrow\mv{\tau} \hat v$.
Assume, without loss of generality, that $v_1 \Rightarrow\mv{\tau} \hat v$.
By induction, $w_1 \Rightarrow\mv{\tau} \hat w$ for a term $\hat w$ without
traces $ax$ or $a^mb$. Now $w \Rightarrow\mv{\tau} \hat w$.

\medskip

\item Suppose $E\vdash v \preccurlyeq w$ because $v=cv_1$ and $w=cw_1$ with
$c\in A$ and $E\vdash v_1\approx w_1$. In this case, proviso (3) of
the lemma can not be met.

\medskip

\item Suppose $E\vdash v \preccurlyeq w$ because $v=\tau v_1$ and
$w=\tau w_1$ with $E\vdash v_1\approx w_1$. Then either $v_1=\hat
v$ or $v_1 \Rightarrow\mv{\tau} \hat v$. In the first case, $w_1$
has no traces $ax$ or $a^{m}b$ by Lem.~\ref{trace-preservation}
and proviso (3) of the lemma; hence $w$ has no such traces either.
In the second case, by induction, $w_1\Rightarrow\mv{\tau} \hat w$
for a term $\hat w$ without traces $ax$ or $a^mb$. Again
$w\Rightarrow\mv{\tau} \hat w$. \qed
\end{itemize}
\end{proof}

\begin{theorem} \label{thm:alphabetn}
If $1<|A|<\infty$, then the inequational theory of $\mathrm{BCCS}(A)$
modulo $\sqsubseteq_{\rm WIF}$ does not have a finite basis.
\end{theorem}

\begin{proof}
Let $E$ be a finite axiomatization over $\mathrm{BCCS}(A)$ that is
sound modulo $\sqsubseteq_{\rm WIF}$. Let $m$ be greater than the
depth of any term in $E$.  According to Lem.~\ref{no-traces2}, the
inequation $\tau(a^mx) + \Phi_m \preccurlyeq \Phi_m$ cannot be derived
from $E$.  Yet it is sound modulo $\sqsubseteq_{\rm WIF}$.
\qed
\end{proof}

\subsubsection{$|A|=1$.} We prove that the inequational theory of
$\mathrm{BCCS}(A)$ modulo $\sqsubseteq_{\rm WIF}$ does \emph{not} have
a finite basis in case of a singleton alphabet. The cornerstone
for this negative result is the following infinite family of
inequations, for $m\geq 0$:
\[ a^m x ~\preccurlyeq~  a^m x + x\]
If $|A|=1$, then these inequations are clearly sound modulo
$\sqsubseteq_{\rm WIF}$. Note that given a closed substitution
$\rho$, $\mc{T}(\rho(x))\subseteq \mc{T}(\rho(a^m x))$.

\begin{lemma}\label{trace-preservation2}
If $t \sqsubseteq_{\rm WIF} u$ then
$\mc{T}_V(t) \subseteq \mc{T}_V(u)$.
\end{lemma}

\begin{proof}
Fix $a\in A$ and an injection $\ulcorner \cdot \urcorner: V
\rightarrow \mathbb{Z}_{>0}$. Let $m=|u|+1$. Define the closed
substitution $\rho$ by $\rho(z)=a^{\ulcorner\!
z\!\urcorner{\cdot}m}\nil$ for all $z\in V$.  By Lem.~\ref{res},
$\mc{CT}(\rho(t)) \subseteq \mc{CT}(\rho(u))$.  Now suppose
$a_1\cdots a_k y \in \mc{T}_V(t)$. Then $a_1\cdots a_k a^{\ulcorner\!
y\!\urcorner{\cdot}m} \in \mc{CT}(\rho(t)) \subseteq
\mc{CT}(\rho(u))$ and $k<m$. This is only possible if $a_1\cdots a_k y
\in \mc{T}_V(u)$. \qed
\end{proof}

\begin{lemma} \label{1}
 Assume that, for terms $t,u$, substitution $\sigma$, action $a$,
 variable $x$, integer $m$:
  \begin{enumerate}\vspace{-1ex}
  \item $t \sqsubseteq_{\rm WIF} u$;

  \item $m>|u|$; and

  \item $x \in \mc{T}_V(\sigma(u))$ and
  $a^k x \not\in \mc{T}_V(\sigma(u))$ for $1\leq k <m$.

\vspace{-1ex}\end{enumerate}
Then $x \in \mc{T}_V(\sigma(t))$ and
$a^k x \not\in \mc{T}_V(\sigma(t))$ for $1\leq k <m$.
\end{lemma}

\begin{proof}

Since $x\in \mc{T}_V(\sigma(u))$, by
Obs.~\ref{traces-substitution} there is a variable $y$ with $y\in
\mc{T}_V(u)$ and $x \in \mc{T}_V(\sigma(y))$. Consider the closed
substitution $\rho$ given by $\rho(y)=a^m\nil$ and $\rho(z)=\nil$
for $z\neq y$. Then $m>|u|=|t|$, and $y\in \mc{T}_V(u)$ implies
$a^m \in \mc{T}(\rho(u)) = \mc{T}(\rho(t))$, so by
Obs.~\ref{traces-substitution} there is some $k < m$ and
$z\mathbin\in V$ such that $a^k z \in \mc{T}_V(t)$ and $a^{m-k}
\in \mc{T}(\rho(z))$. As $k<m$ it must be that $z=y$. Since
$a^ky\in \mc{T}_V(t)$ and $x\in \mc{T}_V(\sigma(y))$,
Obs.~\ref{traces-substitution} implies that $a^k x \in
\mc{T}_V(\sigma(t))$. By Lem.~\ref{trace-preservation2}, $a^k x
\not\in \mc{T}_V(\sigma(t))$ for $1\leq k < m$. Hence we obtain
$k=0$.\qed
\end{proof}

\begin{lemma} \label{lemmaalphabet1}
Assume that, for $E$ an axiomatization sound for $\sqsubseteq_{\rm
WIF}$ and for terms $v,w$, action $a$, variable $x$ and integer $m$:
 \begin{enumerate}\vspace{-1ex}
  \item $E\vdash  v \preccurlyeq w$;

  \item $m > \max \{ |u| \mid t \preccurlyeq u \in  E\}$; and

  \item $x \in \mc{T}_V(w)$ and
  $a^k x \not\in \mc{T}_V(w)$ for $1\leq k < m$.
\vspace{-1ex}\end{enumerate}
Then $x \in \mc{T}_V(v)$ and
$a^k x \not\in \mc{T}_V(v)$ for $1\leq k < m$.
\end{lemma}

\begin{proof}
By induction on the derivation of $E\vdash v \preccurlyeq w$.
\begin{itemize}

\item Suppose $E\vdash v\preccurlyeq w$ because $\sigma(t)=v$ and
$\sigma(u)=w$ for some $t\preccurlyeq u\in E$ and substitution
$\sigma$. The claim then follows by Lem.~\ref{1}.

\medskip

\item Suppose $E\vdash v\preccurlyeq w$ because $E\vdash v\preccurlyeq
u$ and $E\vdash u\preccurlyeq w$ for some $u$. By induction, $x \in
\mc{T}_V(u)$ and $a^k x \not\in \mc{T}_V(u)$ for $1\leq k < m$.
Hence, again by induction, $x \in \mc{T}_V(v)$ and
$a^k x \not\in \mc{T}_V(v)$ for $1\leq k < m$.

\medskip

\item Suppose $E\vdash v\preccurlyeq w$ because $v=v_1+v_2$ and
$w=w_1+w_2$ with $E\vdash v_1 \preccurlyeq w_1$ and $E\vdash
v_2\preccurlyeq w_2$. Since $x \in \mc{T}_V(w)$, either $x \in
\mc{T}_V(w_1)$ or $x \in \mc{T}_V(w_2)$.  Assume, without loss of
generality, that $x \in \mc{T}_V(w_1)$.  Since $a^k x \not\in
\mc{T}_V(w)$ for $1\leq k < m$, surely $a^k x \not\in \mc{T}_V(w_1)$
for $1\leq k < m$.  By induction, $x \in \mc{T}_V(v_1)$, and hence $x
\in \mc{T}_V(v)$.  For $1\leq k < m$ we have $a^k x \not\in
\mc{T}_V(w)$ and hence $a^k x \not\in \mc{T}_V(v)$, by
Lem.~\ref{trace-preservation2}.

\medskip

\item Suppose $E\vdash v \preccurlyeq w$ because $v=cv_1$ and $w=cw_1$ with
$c\in A$ and $E\vdash v_1\approx w_1$. In this case, proviso (3) of
the lemma can not be met.

\medskip

\item Suppose $E\vdash v \preccurlyeq w$ because $v=\tau v_1$ and
$w=\tau w_1$ with $E\vdash v_1\approx w_1$. Then, by proviso (3) of
the lemma, $x \in \mc{T}_V(w_1)$ and $a^k x \not\in \mc{T}_V(w_1)$ for
$1\leq k < m$. By induction, $x \in \mc{T}_V(v_1)$ and $a^k x \not\in
\mc{T}_V(v_1)$ for $1\leq k < m$.  Hence $x \in \mc{T}_V(v)$ and $a^k
x \not\in \mc{T}_V(v)$ for $1\leq k < m$.  \qed
\end{itemize}

\end{proof}

\begin{theorem}\label{thm:alphabet1}
If $|A|=1$, then the inequational theory of $\mathrm{BCCS}(A)$
modulo $\sqsubseteq_{\rm WIF}$ does not have a finite basis.
\end{theorem}

\begin{proof}
Let $E$ be a finite axiomatization over $\mathrm{BCCS}(A)$ that is
sound modulo $\sqsubseteq_{\rm WIF}$. Let $m$ be greater than the
depth of any term in $E$.  According to Lem.~\ref{lemmaalphabet1}, the
inequation $a^mx \preccurlyeq a^m x + x$ cannot be derived
from $E$.  Yet, since $|A|=1$, it is sound modulo $\sqsubseteq_{\rm WIF}$.
\qed
\end{proof}
To conclude this subsection, we have

\begin{theorem}
If $|A|<\infty$, then the inequational theory of $\mathrm{BCCS}(A)$
modulo $\sqsubseteq_{\rm WIF}$ does not have a finite basis.
\end{theorem}
Concluding, in spite of the close resemblance between weak
failures and weak impossible futures semantics, there is a
striking difference between their axiomatizability properties.

\end{document}